\documentclass[sigconf,preprint]{acmart}

\usepackage{algorithmic}
 \usepackage{graphicx}
 \usepackage{textcomp}
 \usepackage{xcolor}

\usepackage{makecell}
\usepackage{multirow}

\usepackage{float}
\usepackage{graphicx}
\usepackage{hyperref}
\usepackage{booktabs}
\usepackage[tight]{subfigure}
\usepackage{caption}

\usepackage{afterpage}

\usepackage{textcomp}
\usepackage{xcolor}
\usepackage{tkz-euclide}
\usepackage{framed}
\usepackage{tikz}
\usepackage{adjustbox}
\usepackage{url}
\usepackage{libertine}
\usepackage{pdflscape}
\usepackage{enumitem}

\usetikzlibrary{positioning, arrows, shapes,chains}
\usepackage{transparent}

\graphicspath{img/}

\usepackage{cleveref}
\Crefformat{figure}{Figure~#1}
\Crefmultiformat{section}{Sections~#2#1#3}{ and~#2#1#3}{, #2#1#3}{ and~#2#1#3}

\AtBeginDocument{%
  }

\copyrightyear{2026}
\acmYear{2026}
\setcopyright{cc}
\setcctype{by-nc-nd}
\acmConference[TechDebt '26]{International Conference on Technical Debt}{April 12--13, 2026}{Rio de Janeiro, Brazil}
\acmBooktitle{International Conference on Technical Debt (TechDebt '26), April 12--13, 2026, Rio de Janeiro, Brazil}
\acmPrice{}
\acmDOI{10.1145/3794915.3795789}
\acmISBN{979-8-4007-2485-5/2026/04}

\begin{document}


\newcommand{\Workshops}{5 }
\newcommand{\WorkshopsText}{five }
\newcommand{\WorkshopsAll}{15 }
\newcommand{\RetrospectivesAll}{4 }

\newcommand{\StudyMonthsAll}{30 }
\newcommand{\ActionCyclesAll}{16 }
\newcommand{\MeetingCntAll}{108 }
\newcommand{\MeetingHrsAll}{96 }
\newcommand{\DoubleObservationAll}{24 }
\newcommand{\ItemsObservedAll}{1132 }
\newcommand{\TDSAGATMeetingCntAll}{28 }
\newcommand{\TDSAGATMeetingAnswersCntAll}{240 }

\newcommand{\ActionCyclesA}{5 }
\newcommand{\ActionCyclesAText}{five }
\newcommand{\RetrospectivesA}{1 }
\newcommand{\StudyMonthsA}{16 }
\newcommand{\StudyStartA}{2023-05 }
\newcommand{\StudyEndA}{2024-08 }

\newcommand{\ActionCyclesB}{6 }
\newcommand{\ActionCyclesBText}{six }
\newcommand{\RetrospectivesB}{2 }
\newcommand{\StudyMonthsB}{16 }
\newcommand{\StudyStartB}{2024-06 }
\newcommand{\StudyEndB}{2025-09 }

\newcommand{\ActionCyclesCText}{five }
\newcommand{\ActionCyclesC}{5 }
\newcommand{\RetrospectivesC}{1 }
\newcommand{\StudyMonthsC}{11 }
\newcommand{\StudyStartC}{2024-12 }
\newcommand{\StudyEndC}{2025-10 }

\newcommand{\MeetingCntA}{40 }
\newcommand{\MeetingHrsA}{24.5 }
\newcommand{\DoubleObservationA}{5 }
\newcommand{\ItemsObservedA}{488 }
\newcommand{\TDSAGATMeetingCntA}{12 }
\newcommand{\TDSAGATMeetingAnswersCntA}{90 }

\newcommand{\MeetingCntB}{57 }
\newcommand{\MeetingHrsB}{53.5 }
\newcommand{\DoubleObservationB}{14 }
\newcommand{\ItemsObservedB} {507 }
\newcommand{\TDSAGATMeetingCntB}{10 }
\newcommand{\TDSAGATMeetingAnswersCntB}{57 }

\newcommand{\MeetingCntC}{11 }
\newcommand{\MeetingHrsC}{18 }
\newcommand{\DoubleObservationC}{5 }
\newcommand{\ItemsObservedC}{137 }
\newcommand{\TDSAGATMeetingCntC}{6 }
\newcommand{\TDSAGATMeetingAnswersCntC}{93 }
\title{A Practical Guide for Establishing a Technical Debt Management Process}


\author{Marion Wiese}
\email{marion.wiese@uni-hamburg.de}
\orcid{0000-0003-0160-9531}
\affiliation{%
  \institution{University of Hamburg}
  \city{Hamburg}
  \country{Germany}
}

\author{Kamila Serwa}
\email{kamila.serwa@uni-hamburg.de}
\orcid{0009-0003-8434-6366}
\affiliation{%
  \institution{University of Hamburg}
  \city{Hamburg}
  \country{Germany}
}

\author{Eva Bittner}
\email{eva.bittner@uni-hamburg.de}
\orcid{0000-0002-7628-6012}

\affiliation{%
  \institution{University of Hamburg}
  \city{Hamburg}
  \country{Germany}
}

\renewcommand{\shortauthors}{Wiese et al.}

\begin{abstract}
\textit{Context.}
Technical Debt (TD) refers to short-term beneficial software solutions that impede future changes, making TD management essential. 
However, establishing a TD management (TDM) process is one of the most pressing concerns in practice. 
\textit{Goal.} 
We plan to identify which previously researched TDM approaches are feasible in practice and what typical challenges emerge to create a guideline for establishing a TDM process.
\textit{Method.} 
We replicated our previously published action research study by conducting five workshops introducing TDM with two teams from different companies. 
To determine the feasibility of TDM approaches, we presented the teams with approaches for various TD activities and let them decide which to adopt.  
Overall, we conducted 19 workshops and retrospectives, analyzing \MeetingCntAll meetings (\MeetingHrsAll hours) over a \StudyMonthsAll-month period. 

\textit{Results.}
The adopted TD prevention strategies and documentation were similar in all teams. 
The teams utilized their respective backlogs and created a new backlog item type for TD, incorporating similar attributes such as \textit{interest}, \textit{contagiousness}, a \textit{resubmission date}, and reminders to discuss drawbacks and risks.
However, they used different prioritization approaches and deviating repayment methods. 
The teams had to overcome similar challenges during the establishment, which we list in this paper.
\textit{Conclusions.} 
We identified the TDM approaches used by all teams as a starting point for best practices. 
For challenges, we provided solutions or identified them as research gaps.
\textit{Issue tracking system} vendors should implement TD issue types employing the identified attributes. 
Finally, we created a white paper for practitioners to establish a TDM process based on our results.

\end{abstract}


\received{20 February 2007}
\received[revised]{12 March 2009}
\received[accepted]{5 June 2009}

\maketitle

    \section{INTRODUCTION}
    \label{sec:Introduction}
        
            Technical debt (TD) is the term for constructs in software systems that are beneficial in the short term, but hinder future changes~\cite{Avgeriou2016a}. 
            These constructs can occur in various elements of the system or software engineering process, e.g., code, tests, or requirements~\cite{Li2015, Ernst2021}. 
            Managing TD is divided into multiple so-called TD activities, such as TD identification, TD documentation, TD measurement, TD prioritization, TD monitoring, TD repayment, and TD prevention~\cite{Li2015}.
            
            Many studies on managing TD have been conducted, and various approaches to managing it have been developed over the last few years~\cite{seaman_2011, DeAlmeida2021, Wiese2022, finke_how_2023, junior_istdm_2023}. 
            Unfortunately, the transfer of these results to practice remains inadequate and was defined as a goal on the research roadmap (Phase 3) of the Dagstuhl Manifesto ``Reframing technical debt''~\cite{avgeriou_manifesto_2025}. 
            Many case studies report the successful establishment of TD management processes and provide valuable approaches~\cite{Yli-Huumo2016,  Guo2016a, Guo2016d, Ramasubbu2019a, Besker2020, DeAlmeida2021, Malakuti2020, Wiese2022, finke_how_2023}. 
            Yet, these studies lack a common strategy and guidance for transferring results to other companies.
            Consequently, a \textit{``shared `standard' best-practice processes and guidelines for TDM''} was also determined as part of the manifesto's roadmap~\cite{avgeriou_manifesto_2025}.
            In our previous study, we proposed a 5-step workshop concept for establishing a TDM process in practice, allowing practitioners to choose from a variety of research approaches.
            Moreover, we offered detailed guidance on how to apply this strategy to other companies~\cite{wiese_establishing_2026}.

            The \textbf{goal} of this study is to create a TDM guide to identify best practices by replicating the workshop approach at TRUMPF SE \& Co. KG and DATEV eG. 
            We aim to uncover new insights by analyzing commonalities and differences in their approaches, and we will assess the challenges they encountered to guide future research. 
            The study addresses the following research questions (RQ):
            
            \begin{itemize}
                
                \item [] 
                \textbf{RQ 1: Which approaches from research do practitioners adopt to manage TD in practice?} 
                By this, we plan to identify commonalities and differences between the companies.
                This may provide initial guidance on establishing best practices for managing TD, which could be further improved through additional replications.
            
                \item [] 
                \textbf{RQ 2: Which challenges commonly emerge when establishing a TD management process in practice?}
                With this RQ, we aim to identify typical challenges that arise during this establishment. The respective solutions identified by the teams support the development of the TDM guide. 
                In cases where the teams encountered challenges but were unable to find a solution, we reveal further research opportunities.
                
                \item [] 
                \textbf{RQ 3: How is the awareness of TD impacted when a process for managing TD is established?}
                Finally, we measure the TD awareness of the participants, as we did in our initial study~\cite{wiese_establishing_2026}. 
                With this, we provide more data on the impact of the 5-step workshop approach and the resulting TDM processes. 
            \end{itemize}

            To answer these questions, we conducted \WorkshopsAll workshops and \RetrospectivesAll retrospectives across all three teams (i.e., including our previous study).
            We analyzed a total of \MeetingCntAll meetings over \MeetingHrsAll hours over a period of \StudyMonthsAll months.          
            We employed systematic observations to analyze the meetings, questionnaires during the workshops, and the so-called TD-SAGAT surveys, interrupting some of the meetings to assess the team's TD awareness, as introduced in our previous paper~\cite{wiese_establishing_2026}.

            As the \textbf{contributions} of our study, we were able to generalize the results of our previous study by identifying TDM approaches used by all teams as a starting point for ``best practices for TDM.'' 
            Those approaches include (1)~establishing a TD champion/manager, (2)~a list of attributes with concrete values to use when documenting TD, (3)~evaluating alternatives with drawbacks and risks to debias decision-making, (3)~prioritizing TD against each other rather than against functional requirements, and (4)~utilizing a \textit{resubmission date}.
            Regarding typical challenges, we identified that teams need (1)~more concrete support than currently given; (2)~a simple approach to identify TD in a backlog; (3)~support on how to start the process; (4)~reminders to adhere to the process in the beginning, preferably by an outside person; (5)~support in formulating a TD item's consequences.
            Based on our results, we developed a white paper for practitioners titled \textit{``A Practical Guide to Establish TDM''}, which is part of our additional material~\cite{AdditionalMaterial}. 
            Additionally, we identified suggestions for \textit{issue tracking system} vendors on how to incorporate TDM into their tools, including a TD issue type for recording TD items in the backlog and visualizing these issues.
            Finally, we were able to support our previous findings that overall awareness of TD increased only when related attributes were included in the issue types as steady reminders. 
            
            
            
            In~\Cref{sec:Method}, we outline our research methodology, which is based on our previous study~\cite{wiese_establishing_2026}. 
            We present the results and answer the research questions in \Cref{sec:Results} and discuss the most relevant insights and suggestions for future research in \Cref{sec:Discussion}. 
            We investigate the related work in \Cref{sec:RelatedWork} and discuss the threats to validity in \Cref{sec:ThreatsToValidity}.
            With \Cref{sec:Conclusion}, we conclude our paper and summarize the contributions for practitioners and researchers.
     
    \section{METHOD}
    \label{sec:Method}
    
    Our study replicates the 5-step workshop approach and employs the action research methodology as presented in our previous study~\cite{wiese_establishing_2026}.
    Below, we provide a brief overview of the action research, introduce the data collection approaches, and the analysis of the data aligned with the RQs.
            

        \subsection{Action Research}
        \label{sec:actionresearch}

        \begin{figure*}
            \centering
            \includegraphics[width=0.9\textwidth]{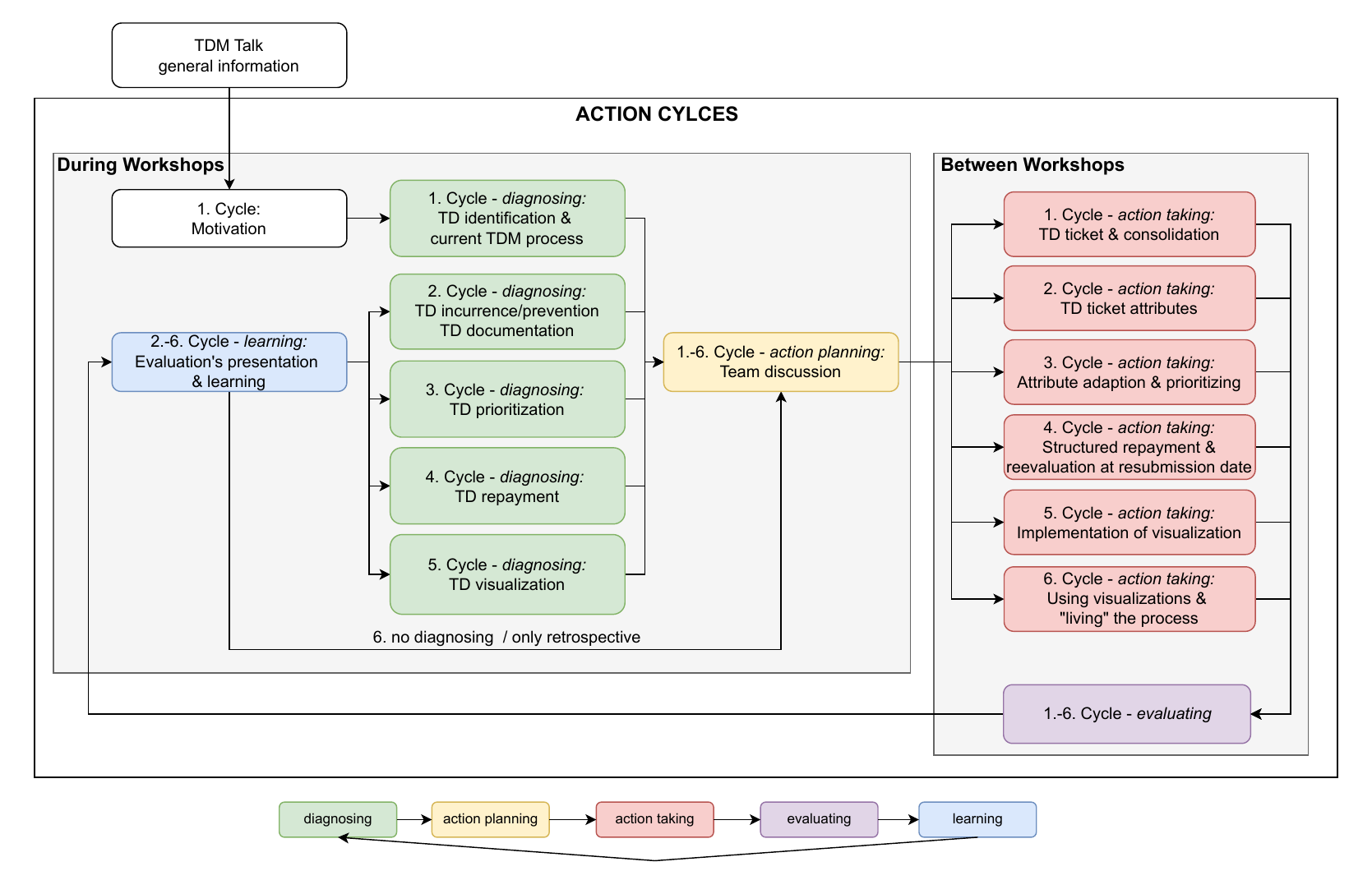}
            \caption{ \centering Study design presenting the action cycles~\cite{wiese_establishing_2026}: \\
            \scriptsize  The workshops are presented in the left grey area and include learning, diagnosing, and action planning. \\  The main actions taken and evaluation between the workshops are shown in the right grey area.}
            \label{fig:Study_Design}
        \end{figure*}
        
            Action research is classified as a field experiment, i.e., it is performed in a natural, pre-existing setting~\cite{Stol2018}.
            In contrast to other research methods, the researchers are allowed to interact with the study's participants~\cite{staron_action_2020}. 
            This offers the advantages of collaborating with practitioners and embedding their feedback directly. 
            These advantages are essential to our goal of understanding why research approaches are adopted or not by practitioners. 
            Moreover, action research follows an iterative approach that aligns with the agile mindset prevalent in most IT teams and our iterative workshop approach. 
            
            Our action research is conducted in iteratively repeating action cycles following Staron's~\cite{staron_action_2020} guidelines. 
            Each action cycle comprises five phases as shown in~\Cref{fig:Study_Design} with different colors for each phase:  
            (1)~\textit{Diagnosing} to identify the problems (green); 
            (2)~\textit{Action planning} to identify actions to solve the problem (yellow); 
            (3)~\textit{Action taking}, i.e., executing the identified actions (red); 
            (4)~\textit{Evaluating} and analyzing the effects and the success of the chosen solutions (violet); 
            (5)~\textit{Learning} to identify further improvements (blue). 
            %
    
        \subsubsection{Case Companies and Participants}
        \label{sec:casecompanies}    
            We replicated our previous study in two teams from different companies with diverse backgrounds and team tasks to offer diverse perspectives. 
            Both companies were selected for this study through the German Software Campus project~\cite{softwarecampus2026}.
            For this paper, we combined our results with those from the initial study to compare the adopted approaches, challenges, and effects. 
            We provide the details on the chosen companies and teams in \Crefrange{tab:companies}{tab:teams} and on the team members in \Cref{tab:participants}.
        
        
    	\begin{table}
    	    \centering
    	    \footnotesize
        	\begin{tabular} { p{2,6cm} p{2cm} p{1,1cm} p{1,8cm} }
    			\toprule
        	   	Company & Domain & Employees  & Organizational Form \\
    			\midrule 
    			DATEV eG	& Finances \& Taxes &   $>$ 8,000 & cooperative \\ 
    			anonymus 	& Consumer goods  & $>$ 2,000  & family business \\ 
    			TRUMPF SE \& Co. KG	& Machine tools & $>$ 15,000 & family business \\ 
     			\bottomrule
    		\end{tabular}
    		\caption{\centering Participating companies, including the one from~\cite{wiese_establishing_2026} \\ \scriptsize (The order of the entries has been randomized for anonymization purposes)} 
            \label{tab:companies}
    	\end{table}

    	\begin{table}
    	    \centering
    	    \footnotesize
        	\begin{tabular} 
                        { p{0,3cm} p{0,3cm} p{4cm} p{2,7cm}} 
    			\toprule
        	   	Team &  Size   & Team's main task & Issue-tracking System \\ 
    			\midrule 
    			A 	&  8 & Providing a central service  & Azure DevOps~\cite{microsoft_azureDevops_2025} \\ 
    			B	&  5 & Frontend (+Backend for Frontend)  & Jira~\cite{Jira2025} \\ 
    			C	&  8 & Central communication platform  & GitLab Issues~\cite{GitLab2025} \\ 
     			\bottomrule
    		\end{tabular}
    		\caption{Participating teams, including team A from~\cite{wiese_establishing_2026}}
            \label{tab:teams}
            \vspace{-0.4cm}
    	\end{table}
        
    	\begin{table}
    	    \centering
    	    \footnotesize
        	\begin{tabular} {ccllcc} 
    			\toprule
        	   	Parti-&  Team & Role & Years  & Gen- & TD \\ 
        	      cipant  &    &   & in IT  & der & manager  \\ 
    			\midrule 
    			P1 	& A & Manager  & 3-5 &  f  &	      \\ 
    			P2 	& A & Software Architect  & $>$10 &  m & x 	\\ 
    			P3 	& A & Engineer  & 6-10 &  m  &	    \\ 
    			P4 	& A & Developer  & 0-2 &  m  & 	    \\ 
    			P5 	& A & Developer  & 6-10 & f &  	    \\ 
    			P6 	& A & Developer  & 15 &  m &  	    \\ 
    			P7 	& A & Developer  & $>$10 &  m &  	    \\ 
    			P8 	& A & Quality Assurance & 0-2 &  m &   \\ 
                
    			P9 	& B & Product Owner   & $>$10 &  m &    \\ 
    			P10 & B & Developer  & $>$10 &  f &  x	    \\ 
    			P11 & B & Developer  & $>$10 &  f &  x   \\ 
    			P12 & B & Developer  & $>$10 &  f &  	    \\ 
    			P13 & B & Developer  & 0-2 &  m &  	    \\ 
                
    			P14 & C & Product Owner  & $>$10 &  m  &	\\ 
    			P15 & C & Software Architect  & $>$10 &  m &  	\\ 
    			P16 & C & Developer  & $>$10 &  m  &	x    \\ 
    			P17 & C & Developer  & 6-10 &  m  &	x    \\ 
    			P18 & C & Developer  & $>$10 &  m  &	    \\ 
    			P19 & C & Developer  & 6-10? &  m  &	    \\ 
    			P20 & C & Developer  & 3-5 &  m  &	    \\ 
     			\bottomrule
    		\end{tabular}
    		\caption{Study Participants including the ones from the initial study of Wiese et al.~\cite{wiese_establishing_2026}}
            \label{tab:participants}
            \vspace{-0.5cm}
    	\end{table}

        \subsubsection{Workshops}
        \label{sec:workshops}
        
            As described in our previous study, each of the five workshops consisted of the \textit{Evaluating}, \textit{Learning}, \textit{Diagnosing}, and \textit{Action Planning} phases of the respective action research cycle.
            The first workshop had a distinct structure from all other workshops.
            The first workshop's focus is on getting to know one another, analyzing the team's current TDM processes, and organizing the upcoming work. 
            Workshops two to five followed a similar structure to introduce TD concepts. 
            First, we presented and discussed the evaluation results, i.e., the results from the surveys 
            and observations (\textit{Evaluating} phase). 
            We employed a retrospective workshop method for the \textit{Learning} phase.
            After that, participants completed a survey to evaluate the study's progress. 
            Then, we presented previously researched approaches regarding one or two TD activities (\textit{Diagnosing} phase).
            The team discussed which presented approaches to employ or developed new approaches to try out (\textit{Action planning} phase).
            
            The second workshop focused on TD documentation, i.e., creating a TD issue type with respective attributes, and TD prevention, i.e., avoiding TD by using quality attributes (QAs) for decision-making and debiasing decision-making by discussing alternatives, drawbacks, and risks. 
            The third workshop focused on TD prioritization factors and a prioritization process.
            The fourth workshop focused on TD repayment strategies and when to use which strategy.
            During the fifth workshop on TD visualization, we discussed questions that should be answered by visualizing the TD issues. 
            An overview of each presented approach is provided in~\Cref{tab:approaches}, including the approaches developed within the initial study~\cite{wiese_establishing_2026} marked with a ``*''.

        \subsubsection{Meetings} 
        \label{sec:meetings}
            Including our first study, we attended \MeetingCntAll meetings, observing the teams for \MeetingHrsAll hours. 
            The teams decided which meetings the researchers were allowed to attend during the kickoff workshop, focusing on meetings involving TD-relevant decisions.
    
            For Team A from our initial study, we participated in the \textit{refinement}, and \textit{review \& planning} meetings, observing \MeetingCntA meetings totaling \MeetingHrsA hours. 
            \textit{Refinement} meetings were used to analyze backlog issues, define an implementation approach, and estimate the effort and other relevant attributes required for each issue. 
            During the \textit{review \& planning} meeting, the team reviewed the success of the former sprint (\textit{review}) and organized its work, i.e., they prioritized the issues for the upcoming sprint (\textit{planning}). 
            Team A recorded the meetings for us to analyze them asynchronously.
            
            For Team B, we participated in the \textit{refinement}, \textit{planning}, and \textit{concept meetings}, observing \MeetingCntB meetings totaling \MeetingHrsB hours.
            The \textit{refinement} and \textit{planning} meetings were similar to the respective meetings from Team A, without reviewing the former sprint. 
            The team further utilized a \textit{concept meeting} to analyze more complex issues. 
            This team did not want their meetings to be recorded, so we had to participate in all meetings synchronously. 
            
            For Team C, we participated in the \textit{refinement} meetings, which were similar to those of Team A and B, observing \MeetingCntC meetings totaling \MeetingHrsC hours. 
            We did not attend the \textit{planning} meetings as the team did not allow us to.
            This team recorded their \textit{refinement} meetings for us. 

        \subsection{Data collection}
        \label{sec:data_collection}
          
    	\begin{table*}
    	    \centering
    	    \footnotesize
        	\begin{tabular} {l|lll|lll|llll|ll} %
    			\toprule
        	   	 &  \multicolumn{6}{c|}{}   &\multicolumn{4}{c|}{Meetings}  & \multicolumn{2}{c}{TD-SAGAT}  \\ 
    			\midrule 
        	   	 & \multicolumn{3}{c|}{Study} & Work- & Retro- & Action &  &  & Discussed  & Observed by  & in $<$x$>$  & No. of  \\ 
        	   	 Team &  Months & Begin & End & shops & spectives & cycles & Cnt. & Hrs &  issues &  2 researchers &  meetings & answers \\ 
    			\midrule 
    			A 	& \StudyMonthsA & \StudyStartA & \StudyEndA & \Workshops & \RetrospectivesA & \ActionCyclesA & \MeetingCntA & \MeetingHrsA & \ItemsObservedA & \DoubleObservationA & \TDSAGATMeetingCntA &  \TDSAGATMeetingAnswersCntA  \\ 
    			B	& \StudyMonthsB & \StudyStartB & \StudyEndB & \Workshops & \RetrospectivesB & \ActionCyclesB &  \MeetingCntB & \MeetingHrsB & \ItemsObservedB & \DoubleObservationB & \TDSAGATMeetingCntB & \TDSAGATMeetingAnswersCntB  \\ 
    			C	& \StudyMonthsC & \StudyStartC & \StudyEndC & \Workshops & \RetrospectivesC & \ActionCyclesC &  \MeetingCntC & \MeetingHrsC & \ItemsObservedC & \DoubleObservationC & \TDSAGATMeetingCntC & \TDSAGATMeetingAnswersCntC  \\  
    			\midrule 
    			Sum & \multicolumn{3}{l|}{\StudyMonthsAll (net 202305--202510; 43 gross)}   & \WorkshopsAll & \RetrospectivesAll & \ActionCyclesAll &  \MeetingCntAll & \MeetingHrsAll & \ItemsObservedAll & \DoubleObservationAll & \TDSAGATMeetingCntAll & \TDSAGATMeetingAnswersCntAll  \\  
     			\bottomrule
    		\end{tabular}
    		\caption{Data collection per Team}
            \label{tab:data_collection}
            \vspace{-0.4cm}
    	\end{table*}
        
            In this section, we present our data collection approaches, which comprise the action research itself, an awareness measurement approach adopted from the psychology domain, our meeting observations, and the self-assessment surveys. 

            \subsubsection{Action Research}
            \label{sec:data_collection_actionreserach}
                All teams conducted \Workshops workshops and at least one retrospective (see~\Cref{tab:data_collection}). 
                We used a study protocol~\cite{AdditionalMaterial} to note down all actions and discussion results for each workshop.
                Additionally, the actions are captured on the last slide of each workshop slide deck~\cite{AdditionalMaterial}, defining process changes and the associated tasks to be performed before the next workshop.
                In addition to workshops and meetings, we communicated with the team via email and chat to clarify specific situations and challenges. 

            \subsubsection{TD-SAGAT -- Situation Awareness Global Assessment Technique}
            \label{sec:data_collection_TDSAGAT}
                \textit{Situation awareness} is a concept from psychology that was proposed by Endsley~\cite{Endsley1988} in 1988.
                As defined by our initial study, we adopted Endsley's definition for TD awareness as follows:
                \textit{TD awareness is the (1)~perception of the prerequisites (e.g., alternatives, benefits, drawbacks) of a decision (e.g., incurring TD) during decision-making situations (e.g., refinement meetings), (2)~the comprehension of these prerequisites' meaning, and (3)~the projection of their status in the future (e.g., TD's consequences).}
                Through the workshops, participants obtain a primary perception of the prerequisites~(1). 
                The action research aims to analyze if participants comprehend those prerequisites~(2) and if they are able to apply them to project future effects~(3) in decision-making situations.
          
                Endsley defined the \textit{Situation Awareness Global Assessment Technique (SAGAT)} to measure \textit{situation awareness}, which we adopted and named TD-SAGAT.
                To identify the relevant prerequisites, the SAGAT approach instructs us to identify the main goals and subgoals for the situation, and the decisions that are necessary to meet the (sub)goal. 
                The situation awareness prerequisites are then derived for these decisions, leading to the queries for a questionnaire by asking whether the participants considered the prerequisites~\cite{Endsley2000}.              
                In our previous study, we identified the following prerequisites for TD decisions:
                (1)~solution alternatives, (2)~solution's benefits, (3)~solution's drawbacks, (4)~solution's risks, (5)~effort for implementing a solution, (6)~TD item's principal, (7)~TD item's interest, (8)~component affected by TD, (9)~functional requirement potentially competing with a TD item, and (10)~quality attributes affected by TD.
                During this study, we recognized that the ninth prerequisite 
                led to multiple follow-up questions and was often not understood.
                We removed this prerequisite from our analysis as the data validity was impacted by these possible misunderstandings.
                
                The idea behind SAGAT is to interrupt participants in their work or a simulation of their work and ask them questions about their situation~\cite{Endsley2000}.
                We employed the TD-SAGAT approach and interrupted the participants at two to three randomly selected time points, approximately every two to three meetings. 
                Then, we asked the participants to complete the TD-SAGAT questionnaire, in which they were asked whether they considered the (randomly ordered) prerequisites during the previous issue's discussion with `yes' or `no' as answer options (see~\Cref{tab:questions}).
                Overall, we used the TD-SAGAT interruptions during \TDSAGATMeetingCntAll meetings, gaining \TDSAGATMeetingAnswersCntAll datasets (see~\Cref{tab:data_collection}). 

            \subsubsection{Observations}
            \label{sec:data_collection_Observations}
            
                We used the same structured observation protocol as proposed in our initial study~\cite {wiese_establishing_2026} for all teams.
                As the meetings were structured around the backlog issues, we recorded our observations per issue, capturing:
                (1) basic information such as date, issue ID, and title, (2) the issue type (TD or not TD) as recorded by the team, as well as our assessment, (3) whether the researchers identified a risk for potential TD incurrence, (4) whether the TD-SAGAT prerequisites were mentioned, and (5) remarks. 
                As remarks, we recorded why we assessed differences for (2) and the details for (3), as well as noticeable events we wanted to discuss with the participants during the upcoming workshop. 
                Of all meetings, \DoubleObservationAll were observed by two researchers who later discussed differences in their assessments to debias the results.
                \Cref{tab:data_collection} provides an overview, per Team, of the study duration, workshops/action cycles, meetings/observations, and TD-SAGAT interruptions.
        
            \subsubsection{Workshop Questionnaires}
            \label{sec:data_collection_questionnaires}

                We created two separate questionnaires: one for the kickoff and another for the subsequent workshops. 
                The kickoff questionnaire differed from the other questionnaires, as we collected basic information on the participants and the current TDM process of the teams. 
                In contrast, the follow-up questionnaires included questions to identify the termination point of the action research, i.e., about the participants' satisfaction with the backlog and with the TDM process.
                Finally, we solicited feedback on the workshops and the study development to refine the workshop approach.
                To measure TD awareness, all questionnaires included nine questions about the TD-SAGAT requirements.
                For this, we asked participants if they usually considered those requirements during decision-making since the last workshop, in a `yes' or `no'  format.
                This is less reliable than a direct TD-SAGAT intervention, but provides us with the participants' self-assessment for method triangulation.
                For space reasons, the complete list of all questions, their types, and purpose is only offered in the initial study~\cite{wiese_establishing_2026} and our additional material~\cite{AdditionalMaterial}. 
                
                

            \begin{table*}[t]
        \centering
        \footnotesize
        \begin{tabular} {p{0.3cm} p{3cm} p{11,8cm} p{1,4cm}  }
            \hline
            \textbf{No.} & \textbf{Question} & \textbf{Sub-Question or answer-options if applicable} & \textbf{Answer type} \\
            \hline
            1  &   \multirow[t]{10}{3cm}{Regarding the topic currently under discussion, I considered ... (Please answer spontaneously! If you have to think long and hard about whether you have (just) thought about something, the answer is probably -- No)}  
                   &  what alternatives are there to the solutions discussed?
            & y/n  \\  
            2  &  & what are the advantages of the solutions discussed?
            & y/n \\  
            3  &  & what are the disadvantages of the solutions discussed?
            & y/n  \\  
            4  & & what are the risks of the solutions discussed?
            & y/n  \\  
            5  &  & how much effort is required for the solutions discussed?
            & y/n  \\  
            6  &  & how much effort (if any) will it take to convert the discussed solutions into the ideal solution (repayment of TS)?
            & y/n  \\  
            7  &  & how high, if any, is the (additional) effort required to permanently maintain the solutions discussed? (Interest from TS)
            & y/n  \\  
            8  &  & which components are affected by the solutions.
            & y/n \\  
            9  &  & which quality attributes are influenced.
            & y/n  \\  
        \end{tabular}
        \caption{\centering Questions from the TD-SAGAT questionnaire, and their answer types. \\
        Similar questions were element of the kickoff and progress questionnaire which are outlined in~\cite{AdditionalMaterial}) }
        \label{tab:questions}
    \end{table*}
        \subsection{Data Analysis}
        \label{sec:data_analysis}
            We present the data analysis aligned with our RQs.
            

            \subsubsection{Adopted Approaches (RQ1)}
            \label{sec:data_analysis_RQ1}

                The used TDM approaches are directly derived from the actions taken between the workshops or retrospectives.
                We focus our evaluation on the final TDM process of each team and do not discuss the course of adopting and dismissing approaches.
                We created an Excel sheet based on the notes in our study protocol, the workshop slides, and our observation protocols.
                We noted down all approaches and indicated for each team whether it was still in use by the end of the study (see~\cite{AdditionalMaterial}). 
                We added remarks to the adoption, e.g., to clarify how the approach was adopted or if the team wished to adopt the approach but was unable to do so due to technical problems.
                
            \subsubsection{Emerging Challenges (RQ2)}
            \label{sec:data_analysis_RQ2}    
            
                We identified and discussed challenges in four different situations:
                First, we observed the participants during their meetings, utilizing an observation protocol as explained above. 
                Second, for some challenges, we used a chat to directly discuss the challenges with the team and receive instant feedback. 
                Third, we used the feedback from the workshop surveys to identify challenges. 
                Fourth, we observed challenges discussed during the workshops.
                From the collected data, we created a final overview in the form of an Excel sheet (see~\cite{AdditionalMaterial}), which describes each challenge, the team affected by this challenge, and how and whether it was resolved.
                In this paper, we focus on the challenges that arose in several teams or that led to valuable solutions.

            \subsubsection{Effects on TD Awareness (RQ3)}
            \label{sec:data_analysis_RQ3}

                To analyze the teams' TD awareness, we (1)~observed the team's discussions (see~\Cref{sec:data_collection_Observations}), (2)~gathered more data with the TD-SAGAT survey~\cite{wiese_establishing_2026} (\Cref{sec:data_collection_TDSAGAT}), and (3)~collected the participants' self-assessments through surveys completed during the workshops (see~\Cref{sec:data_collection_questionnaires}).
                With all three methods, we collected the same data points by asking or looking for the TD requirements as derived from the TD-SAGAT method.
            
                Based on the observation protocols, we counted the number of backlog issues for which the respective prerequisite was discussed per meeting (\textit{observation count}) and the number of all backlog issues (\textit{backlog issues count}).
                We calculated the percentage by dividing the \textit{observation count} by the \textit{backlog issue count}.
                Regarding the TD-SAGAT results, we counted for each prerequisite how often it was considered by any of the participants. 
                This means we summarized their `yes' answers across all participants and interruptions (i.e., backlog issues) per meeting (\textit{TD-SAGAT count}). We further counted the number of all answers we got in a meeting (\textit{meeting's issue count}). 
                We then calculated the percentage by dividing the \textit{TD-SAGAT count} by the \textit{meeting's issue count}.
                From the questionnaire results, we counted the `yes' answers for each prerequisite in the workshop surveys (\textit{survey count}) and the number of all answers per workshop (\textit{workshop answers}).
                For each prerequisite, we calculated the percentage by dividing the \textit{survey count} by the \textit{workshop answers}.

                The three action research studies were performed at different points in time for each team. 
                To build a common timeline across all teams, we had to focus on the results between the respective workshops/retrospectives. 
                For observations and TD-SAGAT, we summarized the values of all meetings between two respective workshops.
                For the workshop questionnaires, the participants assessed the period from the previous action cycle.
                As we also conducted a survey during the kick-off workshop, we gathered survey data for the period preceding the observations and the start of TD-SAGAT, leading to a period ``0. - 1. Workshop'' in the presentation of the results.
                By creating a common timeline, we were able to determine whether the overall awareness of each prerequisite increased during the action cycles.
                



    \section{RESULTS}
    \label{sec:Results}
    
    In contrast to the previous study, this paper focuses on identifying differences and commonalities between the teams to distinguish transferable from context-specific approaches. 
    We present the approaches in~\Cref{sec:Results-approaches}.
    \Cref{sec:Results-challenges} presents common challenges across companies to inform future research directions.
    \Cref{sec:Results-awareness} provides further insights into the development of TD awareness when establishing a TDM process and supports initial research findings.
    Cref{sec:Results-awareness} provides insights into developing TD awareness by establishing a TDM process and supports initial research findings.

        \begin{table*} 
        \centering
        \footnotesize
        \begin{tabular} {p{2cm}  p{7,5cm} p{0.15cm}  p{0.15cm}  p{0.15cm}  p{5,2cm}  }
            \hline 
            \small \textbf{TD activity}  & \small\textbf{Presented approaches} &  \multicolumn{3}{c}{\small\textbf{Adopted by}}   &  \small\textbf{Remark}  \\
                         &                 & \small\textbf{A}  & \small\textbf{B}     & \small\textbf{C} &  \\
            \hline             
            (1) TD communication        
                & \textbf{establishing a TD champion}~\cite{jaspan_defining_2023} 
                    & y & y & y & A+B: Called \textit{TD manager}\\
            \hline    
           (2) TD identification \& definition       
                & Automatically by static analysis tools            
                    & (y) & (y) & (y) &  Only as supporting method \\ 
                & Manually by Dagstuhl definition~\cite{Avgeriou2016a}
                    & (y) & n & n & A: But other TD causes exist \\ 
                & Manually by Kruchten et al.'s backlog colors~\cite{Kruchten2012a} 
                    & n & y & n &  A+C: No external customers exist \\ 
                & \textbf{Manually by identifying whether TD interest occurs}~\cite{avgeriou_technical_2023}
                    & y & y & y &  \\ 
                & \textbf{Manually by identifying who is willing to pay }
                    & y & y & y & \textbf{New approach} \\ 
                    
            \hline 
            (3) TD documentation
                & \textbf{Recording TD in the existing backlog}~\cite{Kruchten2019, Junior2022}   
                    & y  & y & y & \\ 
                & \textbf{Consequences of TD}~\cite{Avgeriou2016a,wiese_it_2023}          
                    & y  & y & y & \\ 
                & *Consequences of repaying TD/``breaking change'' check box~\cite{wiese_establishing_2026}   
                    & y  & y & n & \\ 
                & \textbf{Interest, interest probability}~\cite{Tom2013b, McConnell2008a, Schmid2013}      
                    & y  & y & y & \\ 
                & \textbf{Principal (effort)}~\cite{Tom2013b, McConnell2008a}      
                    & y  & y & y & A: In person days (PD), B: In PD and story points (SP), C: in SP\\ 
                & \textbf{Contagiousness}~\cite{Martini2015a}             
                    & y  & y & y & \\ 
                & \textbf{Alternative, drawbacks, risks} ~\cite{Borowa2021, borowa_debiasing_2022}         
                    & y  & y & y & \\ 
                & Affected quality attributes (QAs)~\cite{iso_25010_2023, bass2012software}   
                    & n  & y & n & \\ 
                & *\textbf{Resubmission date}~\cite{wiese_establishing_2026}           
                    & y  & y & y &  \\ 
                & ``Pain factor'' for the developers (annoyingness of the TD item)      
                    & n  & n & y & \textbf{New approach} \\ 
            \hline 
            (4) TD prevention
                & \textbf{Well-defined} requirements, code \textbf{stan\-dardization,} \-well-\-defined project processes, well-defined architecture/design, using \textbf{good practices}, testing, and reviews \cite{Rios2020, Freire2020a, Perez2021a}
                    & (y)  & (y) &  (y) & The teams generally optimized their work processes; not all strategies were followed\\ 
                & Educational sessions on how to avoid TD \cite{Besker2020} 
                    & n  & n & n & No other sessions than workshops planned\\ 
                & Incorporating TD repayment into project management~\cite{Wiese2022} 
                    & n & n &  n & \\ 
                & \textbf{Debiasing decision-making} by listing at least one solution alternative, one drawback, and one risk per alternative~\cite{Borowa2021, borowa_debiasing_2022} 
                    & y  & y & y & A+C: Additional attribute fields; B: Definition of Ready (DoR)\\ 
                &  Better decision-making by considering QAs~\cite{iso_25010_2023, bass2012software} 
                    & n & y & n & \\ 
                &  Counteract TD inheritance between mechatronic disciplines~\cite{Dong2019, Vogel-Heuser2021}
                    & n  & n & n & \\ 
                &  *``Talked about TD'' checkbox~\cite{wiese_establishing_2026}
                    & y  & (y) & (y) & B: Part of DoR; C: Changed to: \textit{Is it TD?} and \textit{Does it incur TD?}\\ 
            \hline 
            (5) TD measurement 	              
                & \textbf{Using costs (interest, interest probability, principal) for prioritizing}~\cite{Tom2013b, McConnell2008a, Schmid2013} 
                    & y  & y & (y)& To identify \textit{low-hanging fruits} \\ 
                & Using costs (interest, i.probability, principal) for decision-making~\cite{perera_systematic_2024} 
                    & y  & n & n &  \\
                & *Calculating a return on investment (ROI) based on cost attributes~\cite{wiese_establishing_2026}             
                    & (y)  & y & n & A: To compare with an optimize the educated guess\\
                & *Estimating an educated guess based on all attributes~\cite{wiese_establishing_2026}           & y  & (y) & n & B: Attributes like QAs or resource availability might overrule ROI \\ 
                & Mean attribute priority (all attributes on a 1-5 scale)         
                    & n  & n & y & \textbf{New approach}\\ 
            \hline 
            (6) TD prioritiza-             
                & Prioritizing TD against functional requirements~\cite{lenarduzzi_systematic_2020}       
                    & n  & n & n & \\ 
            tion              	         
                & \textbf{Prioritizing TD against each other}~\cite{lenarduzzi_systematic_2020}  
                    & y  & y & y & \\ 
                & Prioritizing once~\cite{lenarduzzi_systematic_2020}          
                    & n & n & n & \\ 
                & \textbf{Prioritizing constantly}~\cite{lenarduzzi_systematic_2020}         
                    & y  & y & y & By using a \textit{re-submission date}\\ 
                & \textbf{Based on the systems evolution}~\cite{Schmid2013}       
                    & y  & y & (y) & C: Not possible due to tool limitations\\ 
                & Based on the business value~\cite{ReboucasDeAlmeida2018c}              
                    & n  & n & n & \\ 
                & \textbf{Based on cost attributes} (TD measurements)~\cite{Tom2013b}             & y  & y & y & A+B: To identify \textit{low-hanging fruits}; C: Included in mean calculation\\ 
                & Based on the affected QAs~\cite{rachow_architecture_2022}        & n & y & n & B: As additional factor\\ 
            \hline 
            (7) TD repay-            
                & \textbf{Repay TD if it impedes a functional requirement}~\cite{FREIRE2023}       & y  & y & y & similar to refactor if it impedes development\\ 
            ment      & \textbf{Pay the interest}~\cite{Buschmann2011}              
                    & y  & y &  y & Use \textit{re-submission date} to recall TD items\\ 
                & \textbf{Wait for discontinuing or migrating} of the affected software system~\cite{Schmid2013, Schmid2013a}              
                    & y  & y & y & Use \textit{re-submission date} to recall TD items \\ 
                & \textbf{Repay TD if the positive effect is proven }by analyzing the costs~\cite{McConnell2008a, Tom2013b}               
                    & y  & y & (y) & repay \textit{low-hanging fruits} if time allows; C: Not possible due to tool limitations \\ 
                & Repay TD by assigning a quota~\cite{McConnell2008a, Wiese2022}              
                    & n  & y & n & \\ 
                & \textbf{Repay if TD might hinder the system's evolution} in the near future~\cite{Schmid2013}              
                    & y  & y & (y)& C: Not possible due to tool limitations\\ 
                & Repay if TD was incurred due to a project deadline~\cite{Wiese2022}               & n  & n & y & Team C planned to use it, but a fitting situation didn't emerge during the study time \\
            \hline 
            (8) TD monitoring \& visualiza-
                & DebtFlag~\cite{holvitie_debtflag_2013}, MultiDimEr~\cite{silva_multidimer_2022}, VisminerTD~\cite{Mendes2019}, MIND~\cite{falessi_towards_2015}, TEDMA~\cite{fernandez-sanchez_open_2017}
                    & n & n & n & Not for backlog visualization, or only for self-developed tools \\   
            tion              	         
                & \textbf{Self-developed, externalized TD visualization (e.g., Microsoft PowerBI) }   
                        & y  & y & (y) & Interfaces to Azure DevOps and Jira exist; C: not possible due to tool limitations \\    
            \hline 
            \hline
        \end{tabular}
        \caption{ \centering Presented approaches with references and their adoption/use by each of the teams in the final TDM process. \\
       \footnotesize (\textbf{bold} -- Approach adopted by all teams; * -- Approach developed in the initial study; (y) -- Approach partly adopted (see remark column)) }
        \label{tab:approaches}
    \end{table*}

        \subsection{Adopted approaches (RQ1)}
        \label{sec:Results-approaches}
        
        In~\Cref{tab:approaches}, we provide an overview of all research approaches presented to the teams, along with their adoption status. 
        The approaches are structured according to the TD activities defined by Li et al.~\cite{Li2015}.
        The remark column clarifies how the approaches were adopted and signals if an approach was newly developed by a team.
        We marked all approaches in bold that were chosen by all teams, and thereby might serve as a starting point for ``best practices.''
        
        The \textit{\textbf{best practices}} we derived from our study and which we present in the TDM Guide are:
        (1) Each team appointed a TD champion/manager to ensure adherence to the process and facilitate communication on TD topics.
        (2) The teams identified TD mostly by considering who ``suffers'' from the TD item and who would pay for its repayment. 
        The teams used automatic TD identification solely as a supporting method. 
        (3) The teams used their backlog, i.e., issue-tracking tool, to tag TD items.
        They created a specific TD issue type or template with various attributes to record their TD items. 
        Common attributes for the TD item are the TD's \textit{interest}, \textit{interest probability}, \textit{principal}, and \textit{contagiousness}. 
        For an ongoing and structured (re-)prioritization, they used a  \textit{resubmission date}.
        The teams measured TD-related attributes differently (e.g., a ``very low'' to ``very high'' scale or a 1 to 5 scale), but provided similar explanations to categorize the genuine values. 
        The teams categorized the  \textit{interest} as: \textit{$<$ 15 min.},  \textit{$<$ 1hr.},  \textit{$<$ 4 hrs.},  \textit{$<$ 8 hrs.}, and\textit{ $>$  8 hrs.} 
        They categorized  \textit{interest probability} as: \textit{once a day}, \textit{once a week}, \textit{once a month}, \textit{once a quarter}, \textit{once a year}.
        The  \textit{contagiousness} was categorized with three values: the principal \textit{decreases}, \textit{remains the same}, or \textit{increases}.
        (4) To avoid TD, the teams added attributes to all backlog item types, reminding them to discuss alternatives, drawbacks, and risks of an implementation approach.
        Various good practices to avoid TD were already part of the team's regular work processes. 
        Some teams refined these processes, e.g., by modifying meeting structures, or utilizing ``Scrum Poker.''
        (5) For measuring the TD items and calculating a prioritization of the TD items, all teams used a different approach: An ``educated guess'' based on all attributes, a ``return on investment'' calculation based on \textit{interest},  \textit{interest probability}, and \textit{effort}, and a ``mean attribute priority'' as a mean of all attributes measured on a one-to-five scale. 
        (6) The teams agreed to prioritize TD against each other for ranking, avoiding competition with functional requirements. 
        They also decided to continuously (re-)prioritize TD items based on the system's evolution using the \textit{resubmission date}.
        (7) For the repayment of TD, all teams adopted multiple approaches depending on the situation. Commonly used approaches include paying the interest (high effort, low priority),  repaying low-hanging-fruits (low effort, high priority), repaying TD items in components that will be changed soon, or combining a TD repayment with the implementation of functional requirements in the same component (system evolution). 
        (8) For visualization purposes, no issue-tracking system provided adequate support. 
        All visualizations of the backlog items' data were externalized to other systems. 
        However, similar visualizations were chosen (see~\cite{AdditionalMaterial}). 

        \begin{table*} 
        \centering
        \footnotesize
        \begin{tabular} { p{0,2cm} p{2,7cm} p{0,1cm} p{0,1cm} p{0,1cm}  p{8,5cm}  p{3,2cm}}
            \hline
                 & \small\textbf{Challenge } &   \multicolumn{3}{c}{\small\textbf{Emerged}}   &  \small\textbf{Challenge description and solution}  &   \small\textbf{Suggestion} \\
            
              & \small\textbf{description}  & \small\textbf{A}  & \small\textbf{B}     & \small\textbf{C}           &                                          & \\
             
            \hline
            \multirow{2}{*}{\raisebox{-3,5\height}{\rotatebox[origin=c]{90}{Identification}}}
            & Getting started: Refining all (old) TD issues 
                & n & y & n & The TD managers and researchers tagged all TD items in the backlog, and teams held extra meetings to refine them.  
                B: Due to the vast number of TD items, they only tagged the items created in the last year. Each team member, then, refined 2-3 issues weekly and set the  \textit{resubmission date} to present them to all developers in subsequent refinements. 
                &  
                Use the Team B method if there are too many open issues in the backlog.\\
            & TD resulting from corporate policy or IT policy forcing changes
                & y & y & y & All teams were unsure how to deal with TD of the company or IT department, which they do not directly suffer from (e.g., switching to a better, company-wide unified issue-tracking tool). 
                Team A + B marked those (mostly) as TD while Team C did not because they did not feel comfortable estimating the cost factors (\textit{interest}, principal, etc.).  
                &  Further research is needed to determine TD items at different stakeholder levels and to explore effective ways to address them. \\    
            & Identifying TD issues
                & y & y & y & Even with an approach of asking who is 'suffering' from the TD item and who would be willing to pay for its repayment, teams were still struggling at times. Team C suggested creating a decision-flowchart and a list of 'TD smells,' indicators in an issue and its discussion that might signify it as a TD item (e.g., specific terms in the title or a developer introducing the issue instead of a product owner). We created both as part of the TDM Guide~\cite{AdditionalMaterial, Wiese_2026_TDMguide}.
                &  Further research should improve the smell list. \\    
            & Remembering to identify TD
                & n & y & y & Team B+C forgot to mark new issues as TD and use the appropriate template. They added a reminder to consider whether an issue is actually a TD item to the template for all issues/DoR.
                &  Adding this reminder to the template from the start.  \\    
                
            \hline
            \multirow{2}{*}{\raisebox{-1,5\height}{\rotatebox[origin=c]{90}{Prevention}}} 
            & Uncovering alternatives
                & n & y & n & Team A+C: The architect led the discussion on alternatives. Team B+C: The team supported the uncovering by changing the discussion order: They started by identifying affected QAs, then derived drawbacks and risks for those QAs to identify the room for improvement, and by this, uncovered alternatives.
                & Further research on supporting the uncovering of solution alternatives on the developer (instead of the architect) level.
                \\   
            &  Debiasing decisions in non-TD items or existing items
                & y  & y & y & We had to (repeatedly) remind the team to assess alternatives (+drawbacks/risks) for all issues (not just TD issues and new issues). Furthermore, we had to instruct the teams to discuss drawbacks and risks even if no alternative existed.
                & Better incorporation into the workshop concept. External support is necessary.  \\          
    
            \hline
            \multirow{5}{*}{\raisebox{-1,7\height}{\rotatebox[origin=c]{90}{Documentation \& Measurement}}}
            & Formulating substan- tial TD consequences 
                & y  & y & y & We had to (repeatedly) explain the method of using ``the five whys''~\cite{serrat_five_2017} to avoid lapidary statements like \textit{`The TD item impacts maintainability.'}, which do not describe the core problem to business stakeholders.
                & Better incorporation into the workshop concept. External support is necessary.  \\     
            & Defining architectural components to use in evolution-based repayment
                & y  & y & y &  We had to (repeatedly) explain the difference between component and cross-cutting concerns to generic components like ``documentation'', ``test'', or ``feature flag removal.'' The evolutionary repayment method can only work if teams are specific when assessing the component. 
                & Better incorporation into the workshop concept. \\          
            & Understanding and using TD-specific terms, esp.  \textit{interest probability} or  \textit{contagiousness}
                & y  & y & y & All teams renamed ``probability'' to ``frequency'' and used \textit{once a week}, \textit{once a month}, etc., as values to better understand the meaning of  \textit{interest probability}. Additionally, all teams explained the fields in the template by adding underlying questions, e.g., ``Will the repayment effort rise over time?'' for  \textit{contagiousness}.
                & Further research on the difference between  \textit{interest probability} and frequency, and how to address both. Researching names that are understood by developers. \\   
            & Addressing multiple \textit{interests} (e.g., additional effort for the team + risk of bugs + security risks).
                & y  & y & y & We specifically discussed this with Team B+C. The teams decided to explain all possible \textit{interests} in the ``Consequences of the TD item'' attribute, but to evaluate only the most critical one in terms of \textit{interest} and  \textit{interest probability}.
                & Further research on dealing with multiple \textit{interests}.\\    
            & Avoiding the central tendency bias for scaled values (e.g., 1-5, very low - very high)
                & y  & n & n & Team A fell for the central tendency bias. Team B+C used categorized real values (e.g., 0-15 Min./15-60 Min./1-4 hrs./4-8 hrs./$>$8hrs.) for  \textit{interest} instead of a ``very low to very high'' scale. As a side effect of this decision, they avoided this bias.
                & Use categorized values at least as an additional explanation. \\  
            \hline  
            \multirow{2}{*}{\raisebox{-1\height}{\rotatebox[origin=c]{90}{Repayment}}} 
            & Integrating a recurring task like feature flag removal or updating 3\textsuperscript{rd} party libraries 
                & n  & y & n & Team B had the issue of regularly updating 3\textsuperscript{rd} party libraries and removing feature flags. Due to the large number of them, they did not want to create TD issues for each library/feature flag. They adopted the procedure: Determine the most pressing removal or update, create an issue, and assess the attributes for this specific item. After this item is repaid, create a new issue for the next most pressing item.
                & Other teams might adopt this approach. \\   
            \hline  
            \multirow{2}{*}{\raisebox{-0,7\height}{\rotatebox[origin=c]{90}{Monitoring \& Visualizing }}}
            & Adapting the issue-tracking tool
                & y  & y & y & All three issue-tracking tools lacked helpful visualization support, necessitating the use of an additional tool for visualization (e.g., Power BI\footnote{\url{https://www.microsoft.com/en-us/power-platform/products/power-bi/}}). The TDM needs more tool maintenance and isn't well-aligned with the team's processes. Using two tools for managing TD raises the chance that the process/visualization won't be utilized effectively. 
                & Issue-tracking tool vendors should support the visualization of issues more effectively. \\    
            & Implementing the required visualization of issues over time (e.g., \textit{interest burden} per time period) 
                & y  & y & y & To identify the number of open issues for the previous time periods (over time), an additional code fragment with a loop over months that evaluates open and closed dates per item is necessary. 
                Since existing tools do not support this, a third component would be required.
                & Issue-tracking tool vendors should support the visualization of issues more effectively. \\   
            \hline
        \end{tabular}
        \caption{Observed challenges, in which team(s) they emerged, explanations or solutions, and suggestions on addressing them} 
        \label{tab:challenges}
    \end{table*}

        \subsection{Emerging Challenges (RQ2)}
        \label{sec:Results-challenges}
        
        We describe the challenges that emerged in all or most of the teams in~\Cref{tab:challenges}.  
        We explain the proposed or adopted resolution we determined and executed during our study for each challenge.
        The overview also explicates suggestions, i.e., improvement measures from the authors' perspectives. 
        The additional material~\cite{AdditionalMaterial} comprises an overview of all challenges, including team-specific ones.
        
        

        \subsection{Effects on TD awareness (RQ3)}
        \label{sec:Results-awareness}

\begin{figure}
        \begin{tabular}{@{}l@{}}
            \subfigure[Survey results per prerequisite and participant (participants answering `yes' versus all participants)]
                {	\label{fig:survey}
                    \includegraphics[width=0.5\textwidth]{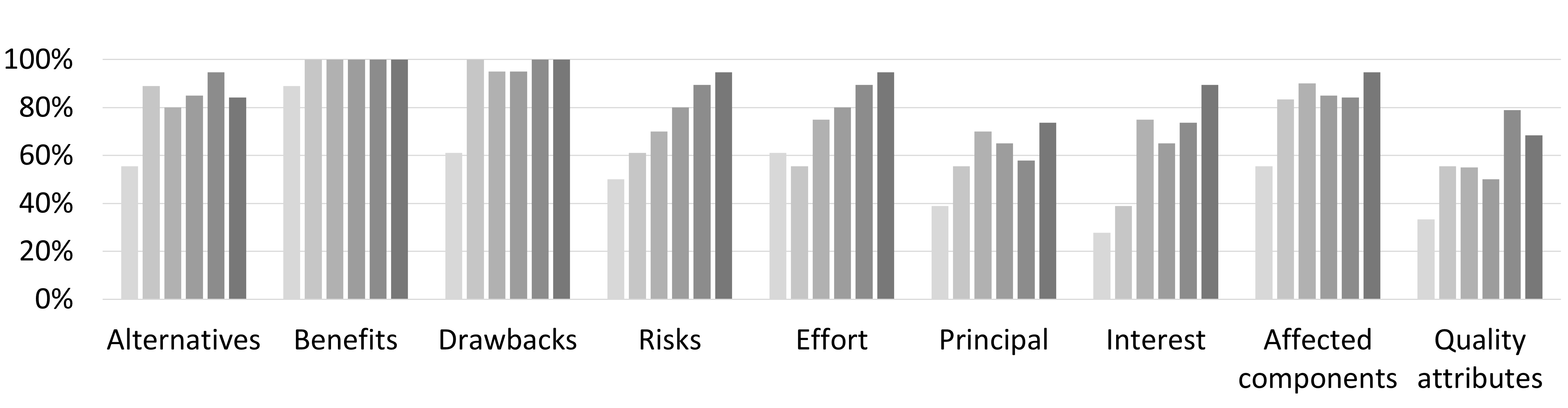}
                } \\
            \subfigure[TD-SAGAT results per prerequisite and answer; (`yes' answers versus all answers)]
                {	\label{fig:TD-SAGAT}
                    \includegraphics[width=0.5\textwidth]{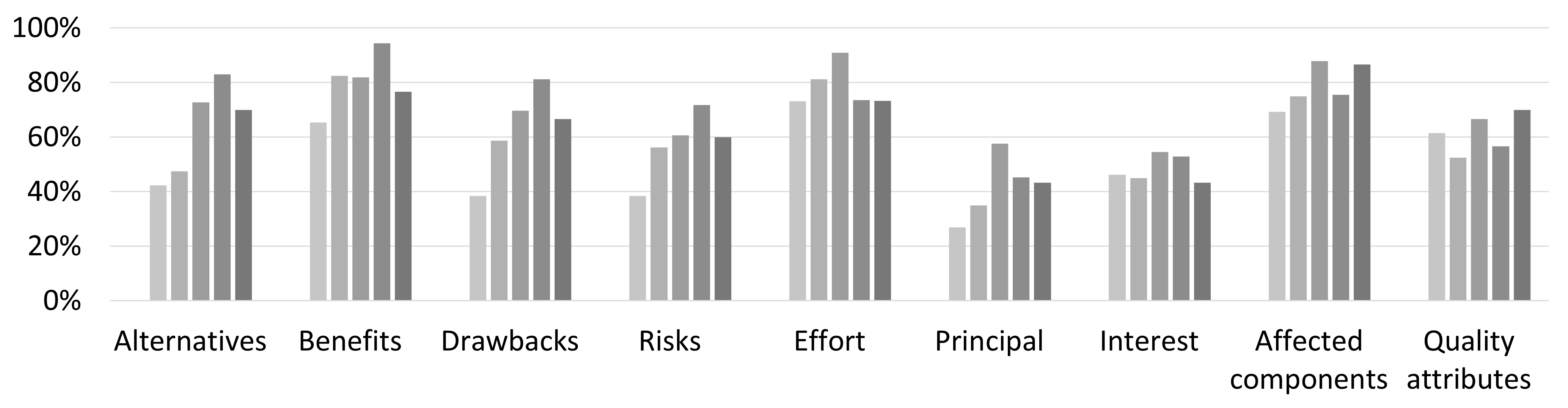}
                } \\
            \subfigure[Observation results per prerequisite (Issues for which a prerequisite was discussed versus all issues (for \textit{interest}, \textit{contagiousness} versus all TD issues))]
                {	\label{fig:Observation}
                    \includegraphics[width=0.5\textwidth]{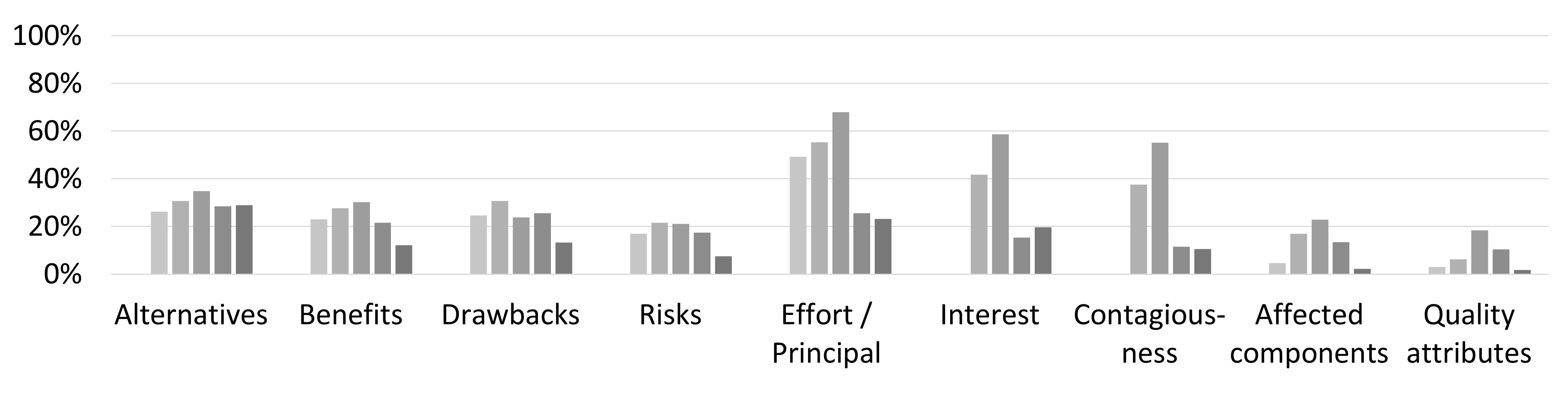}
                } \\
            \subfigure [Legend]
                {	
                    \includegraphics[width=0.5\textwidth]{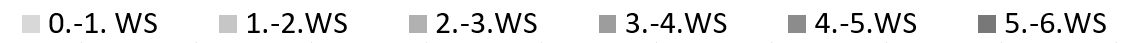}
                }
        \end{tabular}
        \caption{Prerequisites' results measured by a survey, TD-SAGAT surveys, and observations between the workshops (WS)}
        \label{fig:TDawareness}
    \end{figure}

        We present the results of our awareness analysis in~\Cref{fig:TDawareness}.
        The \textbf{self-assessment} in~\Cref{fig:survey} reveals a steady increase in all prerequisites. 
        This is to be expected, as self-assessments are usually less reliable and prone to the bias of ``hypothesis guessing,'' i.e., providing the answer that researchers expect to receive.
        The exception to this rule is the decrease in alternatives between the last workshops, which aligns with the challenge of forgetting to use the template.
        The \textbf{TD-SAGAT results} in~\Cref{fig:TD-SAGAT} also show a steady increase. 
        However, this is on a lower level, indicating that there are situations in which the developers choose not to consider certain attributes. 
        The exceptions to the steady increase are the effort and principal. 
        The effort was initially the highest, and the principal is essentially equivalent to the effort of a TD issue. 
        Considering all other prerequisites may have reduced the focus on this initially most important topic. 
        The \textbf{observation results} in~\Cref{fig:Observation} show that not all prerequisites considered by the participants were actually discussed. 
        Participants stated that they had considered most prerequisites but rarely discussed all of them.
        For the comparison prerequisites (alternatives, benefits, drawbacks, and risks), we observe an expected increase following the second workshop, which included information on cognitive biases. 
        The peak for most of the prerequisites occurred between the third and fourth workshops. 
        Prior to the fourth workshop, the TD issues were primarily refined for the first time in an initial endeavor.
        After this, only a few new TD issues arose, but many were rediscussed due to reaching their \textit{resubmission date}. 
        This means the TD issues were given a new resubmission date, but principal, interest, or contagiousness were not (re-)discussed, leading to a decline in those values.
        Additionally, the decrease after the fourth workshop aligns with the challenge of forgetting to use the template.
        As \textit{affected components} and \textit{quality attributes} were not recorded by all teams, their values tend to be lower than those of the other prerequisites.
        
    \section{DISCUSSION}
    \label{sec:Discussion}
    In this section, we 
    uncover research opportunities, offer support for practitioners, and propose improvements for issue-tracking tools. 

        \subsection{Research}
        \label{sec:Discussion_Reserach}
        
            \paragraph{Concrete support} 
            During our study, we identified multiple instances in which theoretical concepts of TD were too abstract for practitioners to understand or adopt. 
            For example, the concepts of \textit{interest probability} or \textit{contagiousness}, while comprehensible during a talk, were fully understood only when we added a guiding question and renamed \textit{interest probability} to \textit{interest frequency}.
            Another example is the definition of an attribute's content. 
            While the concepts of \textit{interest} and \textit{interest probability} were generally understood, applying them in practice required participants to discuss and define the content of these attributes, i.e., the combo boxes. 
            Overall, we suggest adding more specific support instructions for practitioners, e.g., as enumerated application instructions in the additional material of research papers that claim to benefit practitioners. 
            To adhere to this rule, we added a detailed TDM guide that outlines the TDM process and how to implement it for practitioners to our additional material~\cite{AdditionalMaterial} and published it on Arxiv~\cite{Wiese_2026_TDMguide}.
            
            \paragraph{Research opportunities arising from the challenges}
            We identified research opportunities arising from the challenges we observed (see~\Cref{sec:Results-challenges}).
            We pinpointed the topic of unclear TD identification across multiple stakeholders. 
            In a corporate context, an issue might be a TD item for one stakeholder, but might not be classified as TD from another stakeholder's perspective.
            Following practitioners' suggestions, we created a matrix and a flowchart for TD identification, along with a list of TD indicators, as part of the TDM Guide~\cite{AdditionalMaterial, Wiese_2026_TDMguide}. 
            We further suggested conducting more research on the debiasing of decisions at a lower level of abstraction than architecture. 
            Specifically, we should develop concepts for supporting the identification of alternative solutions. 
            Finally, we encourage the research community to discuss how to prioritize multiple \textit{interests}. 
            
            \paragraph{Key Performance Indicators for Priorizization}
            In this paper, we presented two calculated key performance indicators (KPIs), the ROI and the mean priority value, in addition to the \textit{educated guess} approach used by Team A.
            Because all teams employed different approaches, we were unable to identify a best practice.
            The ROI might enable the generation of helpful visualizations, e.g., the overall \textit{interest burden} per month. 
            This was a specific request from participants from the management perspective. 
            However, tool limitations restricted us from creating such a visualization (see~\Cref{sec:Discussion_Vendors}).
            In contrast, the mean prioritization was easy to use and applicable even when the tool prevented us from creating attribute fields.
            Further research might encompass identifying which KPIs are indeed best practices in terms of context and company-wide aggregation. 

            
        \subsection{Practitioners}
        \label{sec:Discussion_Practitioners}

            \paragraph{Establishing a TDM process}
                Our overall research provides two approaches to establishing a TDM process. 
                First, the workshop concept itself proved to be a valuable approach for establishing a TDM process.
                The primary objective of the workshops is to establish a TDM process iteratively, focusing on each area of the TDM process separately, and allowing for a process tailored to the context.
                %
                %
                Second, we created a white paper outlining a TDM process based on the approaches adopted in this study and solutions to the identified challenges.
                Establishing the process as described in the white paper is less time-intensive and allows for a more standardized approach.

            \paragraph{TD awareness}
            \label{sec:Discussion_awareness}
                As part of our research, we aimed to evaluate whether a TDM process increases overall TD awareness, as measured using the TD-SAGAT prerequisites.
                We demonstrated that adding the attributes to the respective backlog items sustainably increased awareness of single prerequisites.
                Educational workshops raise awareness only immediately after the workshop and can serve only as a first step. 
                This became particularly evident when all teams initially overlooked evaluating alternatives, drawbacks, and risks for non-TD items. 
                Only after adding the attribute fields to the non-TD issue types did the team's awareness of the potential for TD incurrence increase.
                The challenges and decreased values in~\Cref{fig:TDawareness} indicate that a verbal reminder from the researchers seemed necessary at the beginning of the process to internalize its use.
            
            \paragraph{Workshop critique and improvement potential}
                We recognized that the five workshops varied in the complexity of their content. 
                While the second workshop was filled with two different TD activities (documentation \& prevention), the third and fourth workshops were single-focused (prioritization and repayment). 
                As TD prioritization and repayment are closely tied, it may be reasonable to combine both topics in a single workshop to reduce the number of invested person-days to establish the process. 
                Overall, the workshop concept is not yet scalable for big companies. Conducting workshops with all the teams within a company would be expensive. Additionally, the workshop approach would allow each team to develop an individual process, making a cross-company evaluation impossible.
                Consequently, it is essential to identify strategies to enhance the approach's scalability.
                One method would be to conduct the workshops with one or two representative teams, and then allow other teams to adopt the approach, utilizing the developed issue type and visualizations.
                Finally, an alternative would be not to conduct workshops with future teams at all, but to adopt the process as presented in this work (\Cref{sec:Results-approaches} and our guide~\cite{AdditionalMaterial, Wiese_2026_TDMguide}). 
                However, in the questionnaires' open-ended questions, the participants emphasized the value of having an expert present at their refinement meetings: \textit{``Looking at the actual tickets [issues] together helped me the most. If no support is provided at this point, I believe that the greatest added value of the workshops is lost.''}
                Similarly, the evaluation of the TD awareness prerequisites over time demonstrated the need for an external expert.
                Thus, another option is to use a predefined process and refine and establish it with the assistance of an expert. 
            
        \subsection{Tool Vendors}
        \label{sec:Discussion_Vendors}
            In~\Cref{sec:Results-challenges}, we uncovered problems that are specifically related to the tool vendors.
            First, this highlights the importance of selecting an appropriate issue-tracking tool, not only for TDM but also for organizing projects, products, and features. 
            Team C was unable to implement the desired processes and had to endure multiple workarounds, prompting a discussion about employing a more professional issue-tracking tool within the company.

            Second, there is significant potential to enhance professional issue-tracking tools, particularly in their ability to support effective calculations and visualizations.
            While we were able to implement an ROI calculation in Jira, it proved extremely complicated due to the rule-based calculations. 
            Additionally, the standard visualizations in the issue-tracking tools were inadequate for our purpose. 
            In all teams, we had to use an additional visualization tool, which has several disadvantages.
            Besides the effort of implementing and maintaining this additional tool, we observed that teams shied away from or simply forgot to use it during their regular work. 
            
            Finally, some requested visualizations were not feasible even in the additional tools. 
            For example, the participants wanted to visualize the overall \textit{interest burden} (= \textit{interest} * \textit{interest probability}) for all open issues per month to analyze whether the TDM process is indeed successful. 
            We presented the \textit{interest burden} of issues opened or closed at a specific time. 
            To assess the \textit{interest burden} of open issues from a previous month, we needed to evaluate all issues' opened and closed dates. Creating a timeline required looping through all past months and issues, which would have involved developing an additional program.
            We were unwilling to implement this, as our teams could not guarantee the additional long-term maintenance effort. 

    \section{RELATED WORK}
    \label{sec:RelatedWork}

        Two recent studies on TDM by Rios et al.~\cite{rios2018tertiary} and Junior et al.~\cite{Junior2022} provided an overview of TDM research, proving a lack of research on creating a systematic and holistic TDM process. 
        Similarly, the ``Reframing TD manifesto'' developed by Avgeriou et al. identified that TD research should \textit{``manage TD in alignment with its context''}, and that \textit{``a minimum viable TDM process shall be formulated''}~\cite{avgeriou_manifesto_2025}.
        Through our research, we fill this research gap by providing an approach for establishing a context-sensitive TDM process. 
        
        \paragraph{Case Studies and Action Research on TDM}
        Most research on establishing TDM processes in a practical environment has been conducted in the form of case studies~\cite{Yli-Huumo2016, Guo2016d, Ramasubbu2019a, Malakuti2020, Malakuti2020}, which 
        have been tailored to the specific company.
        In contrast, the workshop approach employed in our research does not provide a finalized TDM process but rather a framework for developing one.
        In only a few instances, researchers performed action research~\cite{Oliveira2015, yli-huumo_developing_2016, borup_deliberative_2021, detofeno_technical_2021, detofeno_priortd_2022, wiese_establishing_2026}, which each focused on a few TD activities~\cite{Oliveira2015, yli-huumo_developing_2016, borup_deliberative_2021} or TD types~\cite{detofeno_technical_2021, detofeno_priortd_2022} and did not provide a holistic approach.
        We conducted three action research studies, all following the same structure, which, to the best of the authors' knowledge, has not been previously performed in action research. 

        \paragraph{TD Awareness}
        One separate focus of our study is the analysis of a TDM establishment on TD awareness. 
        While some papers claimed that their TDM approach increased TD awareness, few provided evidence to support this claim~\cite{Gupta2016, Baysal2013, Treude2010, Eliasson2015, Martini2016b, Wiese2022}.
        Researchers typically used self-assessments, i.e., questionnaires~\cite{berenguer_technical_2021, Wiese2022} or concluded TD awareness from indicators derived by static analysis tools~\cite{Crespo2022}.
        To the best of the authors' knowledge, no other study has employed method triangulation to assess TD awareness. 

    \section{THREATS TO VALIDITY}
    \label{sec:ThreatsToValidity}
    	We present threats to validity based on the typical threats to validity for action research provided by Staron~\cite{staron_action_2020}.
    	
        
    	\paragraph{Construct Validity}
            \label{sec:ThreatsToValidity_Construct}
            As we replicated our studies, the ``mono-operation bias'' is mitigated. 
            However, we still only report on three action research projects from one country. 
            The feasibility of the resulting TDM Guide~\cite{AdditionalMaterial,Wiese_2026_TDMguide} in the long run will be revealed by their adoption across various teams. 
            %
            %
            The 'interaction of testing and treatment' is part of the action research's goal, as researchers influence the participants' processes.
            However, the awareness metrics must be evaluated carefully, as our presence in some meetings may have influenced the participants' awareness. 
            Our study may be influenced by ``hypothesis guessing''. 
            To mitigate this, we divided the teams into action and reference teams, emphasizing the value of open criticism at the beginning of each workshop.

    	\paragraph{Internal Validity}

            %
            The voluntary participation leads to a threat of ``biased selection of subject.'' 
            Thus, our results are limited to teams willing to invest in a TDM process. 

    	\paragraph{External Validity}
            Staron describes the "reactive effects of experimental arrangements", which occur when researchers influence the context too much. 
            We have been present in the team's meetings and were sometimes asked about the intricacies of the process, to give more direct feedback, or, on rare occasions, we actively intervened to remind the team of essential process-related commitments. 
            As we see this as the most significant threat, we plan to introduce the process to additional teams using only the TDM guide for future work. 
            %
            In our context, the threat of ``constructs, methods, and confounding factors'' means that the workshop approach is limited to agile working teams, as we relied on team decision-making, limiting the results to teams that are empowered to make such decisions.

    	\paragraph{Conclusion Validity}
            We mitigated the single-observer bias by regularly conducting observations with two researchers, who then discussed their findings together afterward.
            %
            A ``low reliability of measures''  may have influenced the TD-SAGAT results.
            We applied the TD-SAGAT questionnaire only once per month \& team and received about three responses per participant \& meeting.
            However, summarizing the results over all teams remedies this issue to some extent. 

\section{CONCLUSION}
\label{sec:Conclusion}

    In this research, we replicated our five-step workshop approach~\cite{wiese_establishing_2026} in two teams. 
    We identified commonalities and challenges in establishing a TDM process and evaluated changes in the teams' awareness of TD using the TD-SAGAT prerequisites.
    


        \textbf{For practitioners,} we contributed a white paper outlining the establishment of a TDM process. 
        It provides an introduction to the TDM topic, guidelines for the TDM process, a step-by-step guide to establishing the process, a collection of common mistakes and questions with respective suggestions and answers, and an outlook on how to scale such a process at the company level. 
        
        \textbf{For researchers,} we identified various research opportunities relevant to practitioners, including TD identification issues, debiasing decision-making, identifying alternative solutions, and the need for TD-KPI research.
        We urge researchers to offer concrete guidance for practitioners and encourage collaboration with \textit{issue-tracking system} vendors.
    
        \textbf{For Future Work,} we plan to evaluate the TDM guide to identify if a TDM establishment can be performed without external support from researchers or consultants.


\textbf{Acknowledgment and Funding: }
    
    We thank the teams for their participation, trust, and insights.
    The project on which this report is based was sponsored by the Federal Ministry of Research, Technology, and Space of Germany under the funding code 01IS24031. 
    Responsibility for the content of this publication lies with the authors.

\bibliographystyle{ACM-Reference-Format}
\bibliography{TDM_ActionResearch.bib}

\end{document}